\documentclass[aps,prl,twocolumn,showpacs,groupedaddress]{revtex4}  
\usepackage{graphicx}  
\usepackage{dcolumn}   
\usepackage{bm}        
\usepackage{amssymb}   
\newcommand{\MET}   {\mbox{$\not\!\!E_T$}}

\newcommand{\ttbar} {t\bar{t}}

\newcommand{\bbbar} {b\bar{b}}

\def \dzero    {D\O\ }
\begin{document}

\hspace{5.2in} \mbox{Fermilab-Pub-05/207-E}

\title{Search for single top quark production in $p\bar{p}$
collisions at $\sqrt{s}=$1.96~TeV 
}
%
\author{                                                                      
V.M.~Abazov,$^{35}$                                                           
B.~Abbott,$^{72}$                                                             
M.~Abolins,$^{63}$                                                            
B.S.~Acharya,$^{29}$                                                          
M.~Adams,$^{50}$                                                              
T.~Adams,$^{48}$                                                              
M.~Agelou,$^{18}$                                                             
J.-L.~Agram,$^{19}$                                                           
S.H.~Ahn,$^{31}$                                                              
M.~Ahsan,$^{57}$
% ...
\\                                                                            
\vskip 0.30cm                                                                 
\centerline{(D\O\ Collaboration)}                                             
\vskip 0.30cm                                                                 
}                                                                             
\affiliation{                                                                 
\centerline{$^{1}$Universidad de Buenos Aires, Buenos Aires, Argentina}       
\centerline{$^{2}$LAFEX, Centro Brasileiro de Pesquisas F{\'\i}sicas,         
                  Rio de Janeiro, Brazil}                                     
\centerline{$^{3}$Universidade do Estado do Rio de Janeiro,                   
                  Rio de Janeiro, Brazil}                                     
\centerline{$^{4}$Instituto de F\'{\i}sica Te\'orica, Universidade            
                  Estadual Paulista, S\~ao Paulo, Brazil}                     
\centerline{$^{5}$University of Alberta, Edmonton, Alberta, Canada,           
               Simon Fraser University, Burnaby, British Columbia, Canada,}   
\centerline{York University, Toronto, Ontario, Canada, and                    
         McGill University, Montreal, Quebec, Canada}                         
\centerline{$^{6}$Institute of High Energy Physics, Beijing,                  
                  People's Republic of China}                                 
\centerline{$^{7}$University of Science and Technology of China, Hefei,       
                  People's Republic of China}                                 
\centerline{$^{8}$Universidad de los Andes, Bogot\'{a}, Colombia}             
\centerline{$^{9}$Center for Particle Physics, Charles University,            
                  Prague, Czech Republic}                                     
\centerline{$^{10}$Czech Technical University, Prague, Czech Republic}        
\centerline{$^{11}$Center for Particle Physics, Institute of Physics,         
                   Academy of Sciences of the Czech Republic,                 
                   Prague, Czech Republic}                                    
\centerline{$^{12}$Universidad San Francisco de Quito, Quito, Ecuador}        
\centerline{$^{13}$Laboratoire de Physique Corpusculaire, IN2P3-CNRS,         
                  Universit\'e Blaise Pascal, Clermont-Ferrand, France}       
\centerline{$^{14}$Laboratoire de Physique Subatomique et de Cosmologie,      
                  IN2P3-CNRS, Universite de Grenoble 1, Grenoble, France}     
\centerline{$^{15}$CPPM, IN2P3-CNRS, Universit\'e de la M\'editerran\'ee,     
                  Marseille, France}                                          
\centerline{$^{16}$IN2P3-CNRS, Laboratoire de l'Acc\'el\'erateur              
                  Lin\'eaire, Orsay, France}                                  
\centerline{$^{17}$LPNHE, IN2P3-CNRS, Universit\'es Paris VI and VII,         
                  Paris, France}                                              
\centerline{$^{18}$DAPNIA/Service de Physique des Particules, CEA, Saclay,    
                  France}                                                     
\centerline{$^{19}$IReS, IN2P3-CNRS, Universit\'e Louis Pasteur, Strasbourg,  
                France, and Universit\'e de Haute Alsace, Mulhouse, France}   
\centerline{$^{20}$Institut de Physique Nucl\'eaire de Lyon, IN2P3-CNRS,      
                   Universit\'e Claude Bernard, Villeurbanne, France}         
\centerline{$^{21}$III. Physikalisches Institut A, RWTH Aachen,               
                   Aachen, Germany}                                           
\centerline{$^{22}$Physikalisches Institut, Universit{\"a}t Bonn,             
                  Bonn, Germany}                                              
\centerline{$^{23}$Physikalisches Institut, Universit{\"a}t Freiburg,         
                  Freiburg, Germany}                                          
\centerline{$^{24}$Institut f{\"u}r Physik, Universit{\"a}t Mainz,            
                  Mainz, Germany}                                             
\centerline{$^{25}$Ludwig-Maximilians-Universit{\"a}t M{\"u}nchen,            
                   M{\"u}nchen, Germany}                                      
\centerline{$^{26}$Fachbereich Physik, University of Wuppertal,               
                   Wuppertal, Germany}                                        
\centerline{$^{27}$Panjab University, Chandigarh, India}                      
\centerline{$^{28}$Delhi University, Delhi, India}                            
\centerline{$^{29}$Tata Institute of Fundamental Research, Mumbai, India}     
\centerline{$^{30}$University College Dublin, Dublin, Ireland}                
\centerline{$^{31}$Korea Detector Laboratory, Korea University,               
                   Seoul, Korea}                                              
\centerline{$^{32}$CINVESTAV, Mexico City, Mexico}                            
\centerline{$^{33}$FOM-Institute NIKHEF and University of                     
                  Amsterdam/NIKHEF, Amsterdam, The Netherlands}               
\centerline{$^{34}$Radboud University Nijmegen/NIKHEF, Nijmegen, The          
                  Netherlands}                                                
\centerline{$^{35}$Joint Institute for Nuclear Research, Dubna, Russia}       
\centerline{$^{36}$Institute for Theoretical and Experimental Physics,        
                  Moscow, Russia}                                             
\centerline{$^{37}$Moscow State University, Moscow, Russia}                   
\centerline{$^{38}$Institute for High Energy Physics, Protvino, Russia}       
\centerline{$^{39}$Petersburg Nuclear Physics Institute,                      
                   St. Petersburg, Russia}                                    
\centerline{$^{40}$Lund University, Lund, Sweden, Royal Institute of          
                   Technology and Stockholm University, Stockholm,            
                   Sweden, and}                                               
\centerline{Uppsala University, Uppsala, Sweden}                              
\centerline{$^{41}$Lancaster University, Lancaster, United Kingdom}           
\centerline{$^{42}$Imperial College, London, United Kingdom}                  
\centerline{$^{43}$University of Manchester, Manchester, United Kingdom}      
\centerline{$^{44}$University of Arizona, Tucson, Arizona 85721, USA}         
\centerline{$^{45}$Lawrence Berkeley National Laboratory and University of    
                  California, Berkeley, California 94720, USA}                
\centerline{$^{46}$California State University, Fresno, California 93740, USA}
\centerline{$^{47}$University of California, Riverside, California 92521, USA}
\centerline{$^{48}$Florida State University, Tallahassee, Florida 32306, USA} 
\centerline{$^{49}$Fermi National Accelerator Laboratory, Batavia,            
                   Illinois 60510, USA}                                       
\centerline{$^{50}$University of Illinois at Chicago, Chicago,                
                   Illinois 60607, USA}                                       
\centerline{$^{51}$Northern Illinois University, DeKalb, Illinois 60115, USA} 
\centerline{$^{52}$Northwestern University, Evanston, Illinois 60208, USA}    
\centerline{$^{53}$Indiana University, Bloomington, Indiana 47405, USA}       
\centerline{$^{54}$University of Notre Dame, Notre Dame, Indiana 46556, USA}  
\centerline{$^{55}$Iowa State University, Ames, Iowa 50011, USA}              
\centerline{$^{56}$University of Kansas, Lawrence, Kansas 66045, USA}         
\centerline{$^{57}$Kansas State University, Manhattan, Kansas 66506, USA}     
\centerline{$^{58}$Louisiana Tech University, Ruston, Louisiana 71272, USA}   
\centerline{$^{59}$University of Maryland, College Park, Maryland 20742, USA} 
\centerline{$^{60}$Boston University, Boston, Massachusetts 02215, USA}       
\centerline{$^{61}$Northeastern University, Boston, Massachusetts 02115, USA} 
\centerline{$^{62}$University of Michigan, Ann Arbor, Michigan 48109, USA}    
\centerline{$^{63}$Michigan State University, East Lansing, Michigan 48824,   
                   USA}                                                       
\centerline{$^{64}$University of Mississippi, University, Mississippi 38677,  
                   USA}                                                       
\centerline{$^{65}$University of Nebraska, Lincoln, Nebraska 68588, USA}      
\centerline{$^{66}$Princeton University, Princeton, New Jersey 08544, USA}    
\centerline{$^{67}$Columbia University, New York, New York 10027, USA}        
\centerline{$^{68}$University of Rochester, Rochester, New York 14627, USA}   
\centerline{$^{69}$State University of New York, Stony Brook,                 
                   New York 11794, USA}                                       
\centerline{$^{70}$Brookhaven National Laboratory, Upton, New York 11973, USA}
\centerline{$^{71}$Langston University, Langston, Oklahoma 73050, USA}        
\centerline{$^{72}$University of Oklahoma, Norman, Oklahoma 73019, USA}       
\centerline{$^{73}$Brown University, Providence, Rhode Island 02912, USA}     
\centerline{$^{74}$University of Texas, Arlington, Texas 76019, USA}          
\centerline{$^{75}$Southern Methodist University, Dallas, Texas 75275, USA}   
\centerline{$^{76}$Rice University, Houston, Texas 77005, USA}                
\centerline{$^{77}$University of Virginia, Charlottesville, Virginia 22901,   
                   USA}                                                       
\centerline{$^{78}$University of Washington, Seattle, Washington 98195, USA}  
}                                                                             
\date{June 24, 2005}

\begin{abstract}
We present a search for electroweak production of single top quarks in
the $s$-channel 
and $t$-channel 
using neural networks for signal-background separation.  We have analyzed 230~pb$^{-1}$
of data collected with the \dzero detector at the Fermilab Tevatron
Collider at a center-of-mass energy of 1.96~TeV and find no evidence
for a single top quark signal. The resulting 95\% confidence level
upper limits on the single top quark production cross
sections are 6.4~pb in the $s$-channel and 5.0~pb in the $t$-channel.
\end{abstract}

\maketitle

\normalsize

\vfill

\newpage

\begin{widetext}
\end{widetext}

%
%
\section{Introduction}
Top quark physics provides fundamental knowledge of the
strong and electroweak sectors of the standard model and offers discovery
potential for physics beyond the standard model. The
top quark was discovered in 1995 at the Fermilab Tevatron Collider in
$t \bar{t}$ events produced through the strong
interaction~\cite{topdiscovery}. The standard model predicts that
proton-antiproton collisions should also produce single top quarks through
the electroweak interaction. Studying single top quark production will
provide direct measurements of the CKM matrix element $\left| V_{tb} \right|$ and 
top quark polarization, and will probe possible 
new physics in the top quark sector~\cite{Chakraborty:2003iw,Tait:2000sh}. 

\begin{figure}[h!tbp]  
\begin{center}
\includegraphics[width=0.22\textwidth]
{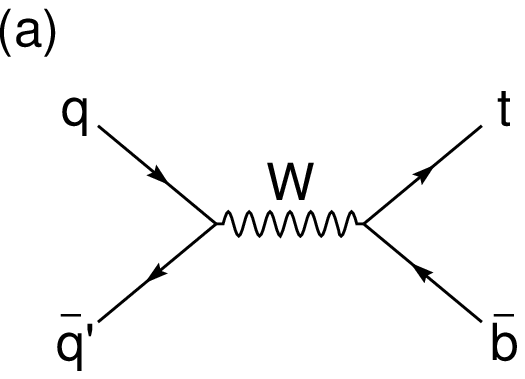}
\includegraphics[width=0.22\textwidth]
{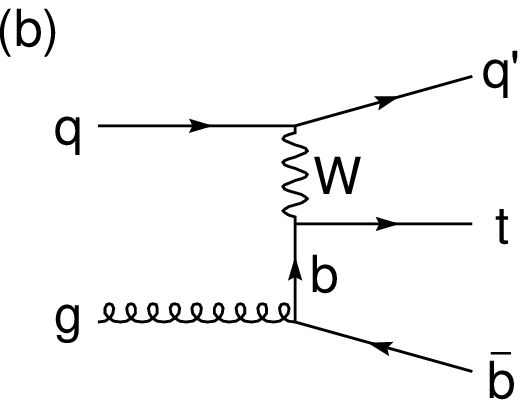}
\end{center}
\caption[tevprod]{Representative Feynman diagrams for electroweak top quark production
at the Tevatron Collider. This figure shows (a) the $s$-channel
and (b) the $t$-channel.}
\label{tev_production}
\end{figure}

There are two main modes of single top
quark production as shown in Fig.~\ref{tev_production}: 
the $s$-channel ($tb$) process $p \bar{p} \rightarrow t \bar{b} + X$
and the $t$-channel ($tqb$) process $p \bar{p} \rightarrow t q
\bar{b} + X$. The production cross sections have been calculated
at next-to-leading order (NLO) in the strong
coupling constant~\cite{sintop-xsec1,sintop-xsec2,Campbell:2004ch,sintop-nlo-sch,sintop-nlo-tch}, yielding
$0.88\pm0.14$~pb for the $s$-channel and $1.98\pm0.30$~pb~\cite{sintop-xsec1,sintop-xsec2}
for the $t$-channel, assuming a top quark mass of $m_{t}=175$~GeV.
Both the \dzero and CDF collaborations have previously performed
searches for single top quark production~\cite{d0runI,Acosta:2001un}. 
Recently, CDF performed a search using 160~pb$^{-1}$ of data and obtained upper
limits of 13.6~pb ($s$-channel),
10.1~pb ($t$-channel), and 17.8~pb ($s$+$t$ combined) at 95\% C.L.~\cite{RunII:cdf_result}. 
%
%

In this Letter, we present a new search for electroweak production of single 
top quarks with the \dzero detector. 
Our search focuses on the final state in which the top quark decays
into a $b$~quark  and a $W$~boson, and the $W$~boson then  decays into
an electron or a muon, and a neutrino. This gives rise to an event
signature with a high transverse momentum ($p_{T}$) lepton with
significant missing transverse energy ($\MET$)
from the neutrino. In addition, the $s$-channel has two $b$-quark
jets, whereas the $t$-channel typically has one $b$-quark jet together
with a light-quark jet. 
The $\bar{b}$-quark jet in the $t$-channel, as represented in 
Fig.~\ref{tev_production}(b), is usually emitted in the forward
direction with low $p_{T}$ and is often undetected~\cite{sintop-xsec2,sintop-nlo-tch}. 
The main backgrounds in
this analysis are the $W$~boson production in association with jets ($W$+jets), top
quark pair ($\ttbar$), and multijet production.
We use neural networks to separate the signals from the
backgrounds. In the absence of 
any significant evidence for signal, we build a binned likelihood from 
the neural network outputs to set upper limits on the single top quark
production cross sections.
%
%
\section{The \dzero Detector}
The \dzero detector consists of a central tracking system, 
liquid-argon/uranium calorimeters, and an iron toroid muon
spectrometer~\cite{D0detector}. The central tracking system covers the
detector pseudorapidity~\cite{fiducial_endnote} region $|\eta_{\rm det}|<2.5$.
It includes a silicon microstrip tracker and a scintillating fiber
tracker, both located within a 2~T solenoidal magnet. The calorimeters
consist of a central barrel in the region $|\eta_{\rm det}|<1.1$,
and two end caps extending the
coverage to $|\eta_{\rm det}| \approx 4.2$. The muon system outside
the calorimeter consists of a layer of tracking detectors and
scintillation counters before 1.8~T toroids, followed by two similar
layers after the toroids.  Luminosity is measured using plastic
scintillator arrays located in front of the end calorimeters.

%
%
\section{Data Set and event selection}
The data were recorded
between August 2002 and March 2004 using a
trigger that required an electromagnetic energy cluster and a jet in the
calorimeter for the electron channel, or
a muon and a jet for the muon channel. The integrated luminosity is
226~pb$^{-1}$ for the electron channel and 229~pb$^{-1}$ for the muon channel.

In the electron channel, we require
exactly one isolated electron~\cite{top-cs-topo}
with $p_{T} > 15$~GeV and $|\eta_{\rm det}|< 1.1$. In the muon
channel, events are selected by requiring exactly one isolated muon~\cite{top-cs-topo}
with $p_{T} > 15$~GeV and $|\eta_{\rm det}|<2.0$. For both channels,
events are also required to have
$\MET>15$~GeV. Events must have from two to four jets 
with the leading jet $p_{T}>25$~GeV and $|\eta_{\rm det}|<2.5$, and all
other jets having $p_{T}>15$~GeV and $|\eta_{\rm det}|<3.4$. 
Jets are defined using a cone algorithm with radius 
${\cal{R}} = 0.5$~\cite{jet_def}.
Misreconstructed events are rejected by requiring that the direction
of $\MET$ is not aligned
or anti-aligned in azimuth with the lepton or the jets. 

The fraction of signal-like events is further enhanced through the
selection of $b$-quark jets that are identified by reconstructing
displaced vertices from long-lived particles. 
A displaced vertex is selected by requiring the
transverse decay-length significance, $L_{xy}/\sigma_{L_{xy}}$, to be greater than
seven, where $L_{xy}$ is the decay length and $\sigma_{L_{xy}}$ is the
uncertainty on $L_{xy}$, calculated from the error matrices
of the tracks and the primary vertex. A jet is considered $b$-tagged by this
algorithm if a displaced vertex lies within a cone of radius
${\cal{R}} = 0.5$ around the jet axis~\cite{top-cs-btag}.

For both $s$-channel and $t$-channel searches, we separate the data
into independent analysis sets based on final-state 
lepton flavor (electron or muon) and $b$-tag multiplicity. To take
advantage of the different final state topologies, we separate
single-tagged (=1~tag) events from double-tagged ($\geq$2~tags)
events.
In the $t$-channel search,
we additionally require that one of the jets is not $b$ tagged. 
%
%
\section{Acceptances and Yields}
We estimate the kinematic and geometric acceptances for $s$-channel 
and $t$-channel single top quark production using the {\sc comphep} matrix element
event generator~\cite{comphepref} with $m_{t}=175$~GeV.
The factorization scales are
$m_{t}^2$ for the $s$-channel samples and $(m_{t}/2)^2$ for the $t$-channel samples. 
For the $s$-channel ($t$-channel) search, the $t$-channel ($s$-channel)
is considered as background.

We use both Monte Carlo events and data to estimate the background
yields. The $W$+jets and diboson ($WW$ and $WZ$) backgrounds are
estimated using events generated with {\sc
alpgen}~\cite{Mangano:2002ea}. The diboson background yields are
normalized to NLO cross sections computed with {\sc
mcfm}~\cite{mcfmref}. 
The total $W$+jets yield is normalized to the yield in data corrected
for the presence of multijet, $\ttbar$ and dibosons before requiring a
$b$-tagged jet.
The fraction of heavy-flavor ($Wb\bar{b}$) events is obtained using
the ratio of the NLO cross sections for $W$+jets and $Wb\bar{b}$, as
described in Ref.~\cite{Abazov:2004au}.
This normalization to data also
accounts for smaller contributions such as $Z$+jets events in which one
of the leptons from the $Z$~boson decay is not reconstructed. 

The {$\ttbar$} background, consisting of the leptonic decay modes of the $W$~boson 
from the top quark decay ($\ell$+jets and dilepton),
is estimated using samples generated
with {\sc alpgen}, normalized to the cross section:
$\sigma(\ttbar)=6.7 \pm 1.2$~pb~\cite{Kidonakis:2003qe}, where the
uncertainty on the top quark mass is incorporated into the cross section
uncertainty.

The parton-level samples are then processed with {\sc
pythia}~\cite{pythiaref} for hadronization, particle decays, and
modeling of the underlying event.
The generated events are processed through a 
{\sc geant}-based~\cite{geantref} simulation of the \dzero
detector. The resulting lepton and jet energies are smeared to
reproduce the resolutions observed in data. 

The background from jets misidentified as electrons or jets resulting
in isolated muons is estimated using 
multijet data samples 
that pass all event selection cuts, but fail the requirement on muon 
isolation or electron quality~\cite{top-cs-topo}.
This background is normalized using a data sample dominated by multijet
events, selected by requiring $\MET < 15$~GeV. 

The overall acceptances, including trigger and selection efficiencies,
for signal events with at least one $b$-tagged jet are $(2.7\pm0.2)\%$
in the $s$-channel and $(1.9\pm0.2)\%$ in the $t$-channel.
The acceptance is calculated
as the fraction of events that pass the selection over all possible
single top quark decays, including all leptonic and hadronic decays of the
$W$~boson. Estimates for signal and background yields and the
observed numbers of events after selection are shown in
Table~\ref{tab:yield_bkg_and_sig}.
%
%
\begin{table}[!h!tbp]
\begin{center}
\caption{Estimates for signal and background yields and the numbers
of observed events in data after event selection for the electron and
muon, single-tagged and double-tagged analysis sets combined. The $W$+jets yields
include the diboson backgrounds.  The total background for the $s$-channel
($t$-channel) search includes the $tqb$ ($tb$) yield. The quoted yield
uncertainties include systematic uncertainties taking into account correlations
between the different analysis channels and samples.}
\begin{ruledtabular}
\begin{tabular}{lr@{$\,\pm\!\!\!\!$}lr@{$\,\pm\!\!\!\!$}l} 
Source           & \multicolumn{2}{c}{$s$-channel search} &
\multicolumn{2}{c}{$t$-channel search} \\ \hline
$tb$             &   5.5 & 1.2  &   4.7 & 1.0  \\ 
$tqb$            &   8.6 & 1.9  &   8.5 & 1.9  \\ 
$W$+jets         & 169.1 & 19.2 & 163.9 & 17.8 \\
$\ttbar$         &  78.3 & 17.6 &  75.9 & 17.0 \\
Multijet         &  31.4 & 3.3  &  31.3 & 3.2  \\ \hline
Total background & ~~~287.4 & 31.4 & ~~~275.8 & 31.5 \\
Observed events  & \multicolumn{2}{c}{283}  &  \multicolumn{2}{c}{271}     \\
\end{tabular}
\end{ruledtabular}
\label{tab:yield_bkg_and_sig}
\end{center}
\end{table}
%
%
\section{Neural Networks analysis}
After event selection, several variables are combined
in neural networks to discriminate the single top quark
signals from the backgrounds. The networks are composed of three
layers of nodes: input, hidden, and output. For training and testing, we use the
{\sc mlpfit}~\cite{mlpfit} package. Testing and training event sets are
created from simulated signal and background samples. 
We use a technique called early stopping~\cite{early_stopping} to determine the 
maximum number of epochs for training which prevents over-training.
Each network is then tuned by choosing the
optimal number of hidden nodes. From studies based on optimizing the
expected upper limits on the single top quark production cross sections, we find that 
the $s$-channel and $t$-channel searches each require
only two networks, corresponding
to the dominant backgrounds: $W b \bar{b}$ and $\ttbar
\rightarrow \ell$+jets. 

The list of discriminating variables has been chosen based on an 
analysis of Feynman diagrams of signals and backgrounds~\cite{boos-dudko} 
and on a study of single top quark production at NLO~\cite{sintop-nlo-sch,sintop-nlo-tch}. 
The  input variables to each network are selected from this list
by training with different combinations of variables and choosing 
the combination that produces the
minimum testing error and largest signal-background separation.
Table~\ref{tab:variable-sets} shows the variables used for
each signal-background pair.  These variables 
fall into three categories: individual-object
kinematics, global-event kinematics, and angular correlations.
%
%
\begin{table*}[!h!tbp]
\begin{center}
\caption[tab:variable-sets]{Input variables for each neural network
signal-background pair. Variable descriptions can be found in the text.
}
\begin{footnotesize}
\begin{ruledtabular}
\begin{tabular}{lp{0.60\textwidth}cccc} 
\multicolumn{6}{r}{Signal-Background Pairs} \\
& & \multicolumn{2}{c}{$tb$} &  \multicolumn{2}{c}{$tqb$} \\
\multicolumn{1}{c}{Variable}&\multicolumn{1}{c}{Description} &      $W\bbbar$  & ${\ttbar}$ &  $W\bbbar$ & ${\ttbar}$ \\
\hline
\multicolumn{6}{c}{\bf{Individual object kinematics}} \\
$p_T({\rm jet1}_{\rm tagged})$     & 
Transverse momentum of the leading tagged jet     & $\surd$ & $\surd$ & $\surd$ & --- \\            
$p_{T}({\rm jet1}_{\rm untagged})$ & 
Transverse momentum of the leading untagged jet   & --- & --- & $\surd$ & $\surd$ \\            
$p_{T}({\rm jet2}_{\rm untagged})$ & 
Transverse momentum of the second untagged jet    & --- & --- & --- & $\surd$ \\            
$p_{T}({\rm jet1}_{\rm nonbest})$ & 
Transverse momentum of the leading nonbest jet   & $\surd$ & $\surd$ & --- & --- \\            
$p_{T}({\rm jet2}_{\rm nonbest})$ & 
Transverse momentum of the second nonbest jet    & $\surd$ & $\surd$ & --- & --- \\            
\multicolumn{6}{c}{\bf{Global event kinematics}} \\
$M_T({\rm jet1},{\rm jet2})$        & 
Transverse mass of the two leading jets     & $\surd$ & --- & --- & --- \\
$p_T({\rm jet1},{\rm jet2})$        & 
Transverse momentum of the two leading jets& $\surd$ & --- & $\surd$ & --- \\            
$M({\rm alljets})$           & 
Invariant mass of all jets           & $\surd$ & $\surd$ & $\surd$ & $\surd$ \\ 
$H_T({\rm alljets})$         & 
Sum of the transverse energies of all jets      & --- & --- & $\surd$ & --- \\
$M({\rm alljets}-{\rm jet1}_{\rm tagged})$ & 
Invariant mass of all jets excluding the leading tagged jet   & --- & --- & --- & $\surd$ \\  
$H({\rm alljets}-{\rm jet1}_{\rm tagged})$ & 
Sum of the energies of all jets excluding the leading tagged jet & --- & $\surd$ & --- & $\surd$ \\ 
$H_T({\rm alljets}-{\rm jet1}_{\rm tagged})$ & 
Sum of the transverse energies of all jets excluding the leading tagged jet         & --- & --- & --- & $\surd$ \\ 
$p_T({\rm alljets}-{\rm jet1}_{\rm tagged})$ & 
Transverse momentum of all jets excluding the leading tagged jet    & --- & $\surd$ & --- & $\surd$ \\
$M({\rm alljets - jet_{best}})$ & 
Invariant mass of all jets excluding the best jet          & --- & $\surd$ & --- & --- \\            
$H({\rm alljets}-{\rm jet}_{\rm best})$ & 
Sum of the energies of all jets excluding the best jet    & --- & $\surd$ & --- & --- \\      
$H_T({\rm alljets}-{\rm jet}_{\rm best})$ & 
Sum of the transverse energies of all jets excluding the best jet     & --- & $\surd$ & --- & --- \\
$M(W,{\rm jet1}_{\rm tagged})$ & 
Invariant mass of the reconstructed top quark using the leading tagged jet & $\surd$ & $\surd$ & $\surd$ & $\surd$ \\
$M(W,{\rm jet}_{\rm best})$ & 
Invariant mass of the reconstructed top quark using the best jet  & $\surd$ & --- & --- & --- \\            
$\sqrt{\hat{s}}$ &
Invariant mass of all final state objects 
                                       & $\surd$ & --- & $\surd$ & $\surd$ \\    
\multicolumn{6}{c}{\bf{Angular variables}} \\
$\Delta \cal{R}({\rm jet1},{\rm jet2})$ &
Angular separation between the leading two jets     & $\surd$ & --- & $\surd$ & --- \\ 
$\eta({\rm jet1}_{\rm untagged}) \times Q_{\ell}$ & 
Pseudorapidity of the leading untagged jet  $\times$ lepton charge     & --- & --- & $\surd$ & $\surd$ \\
$\cos({\rm \ell},Q_{\ell}$$\times$$z)_{\rm top_{best}}$ &
Top quark spin correlation in the optimal basis for the $s$-channel~\cite{Mahlon:1995zn}, reconstructing the top quark with the best jet  & $\surd$ & --- & --- & --- \\ 
$\cos({\rm \ell},{\rm jet1}_{\rm untagged})_{\rm top_{tagged}}$      &
Top quark spin correlation in the optimal basis for the $t$-channel~\cite{Mahlon:1995zn}, reconstructing the top quark with the leading tagged jet & --- & --- & $\surd$ & --- \\            
$\cos({\rm alljets},{\rm jet1}_{\rm tagged})_{\rm alljets}$          &            
Cosine of the angle between the leading tagged jet and the alljets system in the alljets rest frame     & --- & --- & $\surd$ & $\surd$ \\ 
$\cos({\rm alljets},{\rm jet}_{\rm nonbest})_{\rm alljets}$  &
Cosine of the angle between the leading non-best jet and the alljets system in the alljets rest frame  & --- & $\surd$ & --- & --- \\ 
\end{tabular}
\end{ruledtabular}
\end{footnotesize}
\label{tab:variable-sets}
\end{center}
\end{table*}
%
%
\begin{figure*}[!h!tbp]  
\begin{center}
\includegraphics[width=0.385\textwidth]
{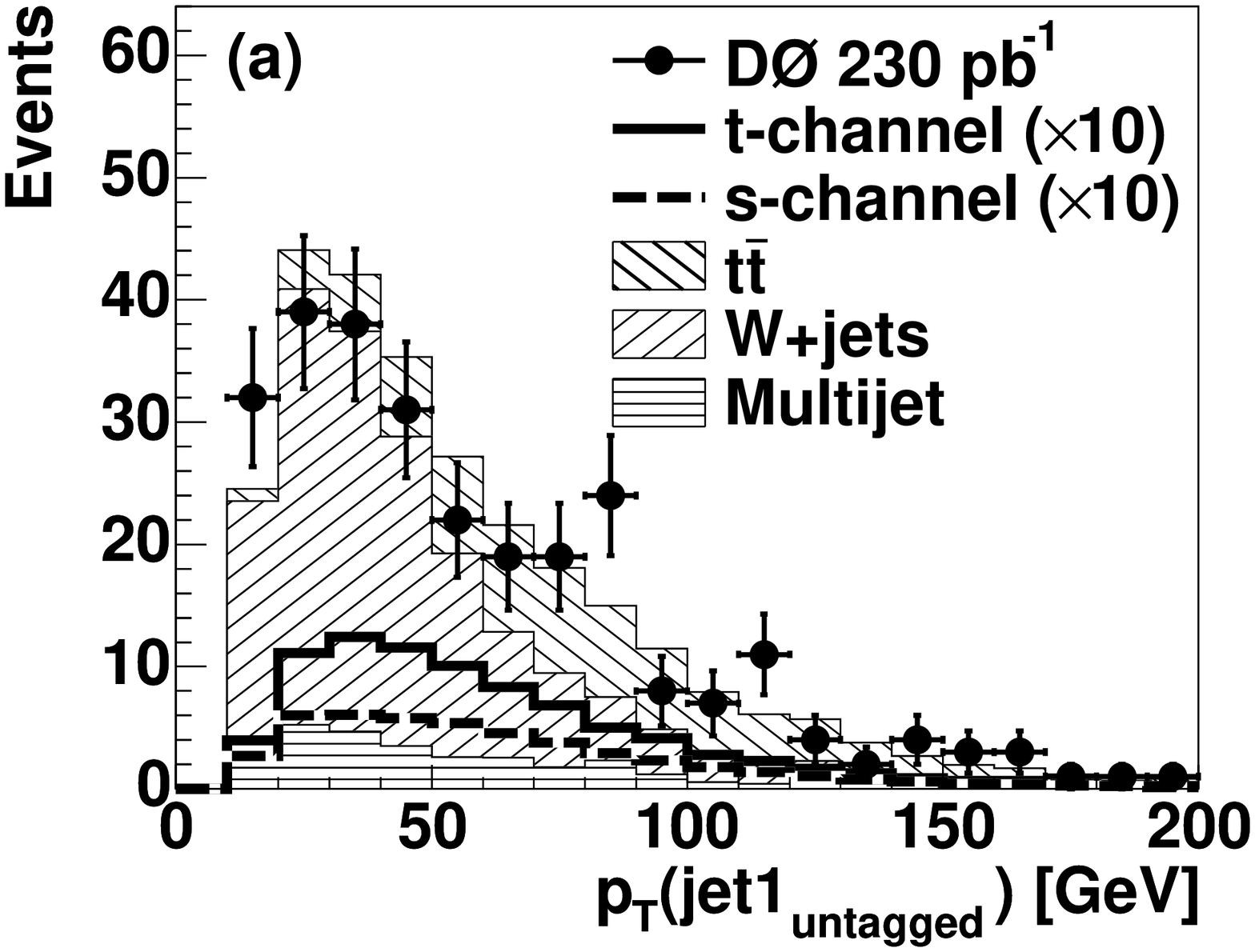}
\includegraphics[width=0.385\textwidth]
{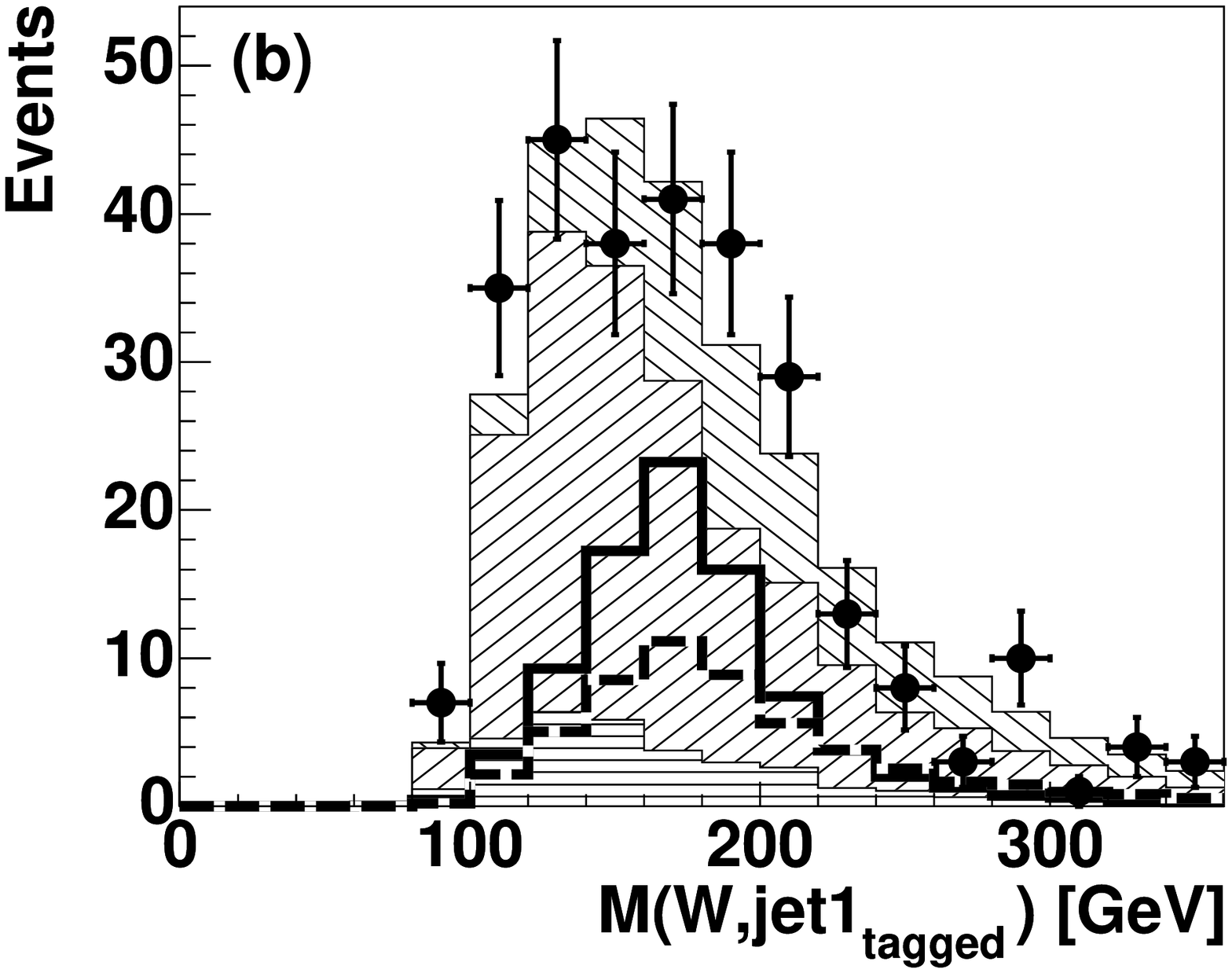}
\includegraphics[width=0.385\textwidth]
{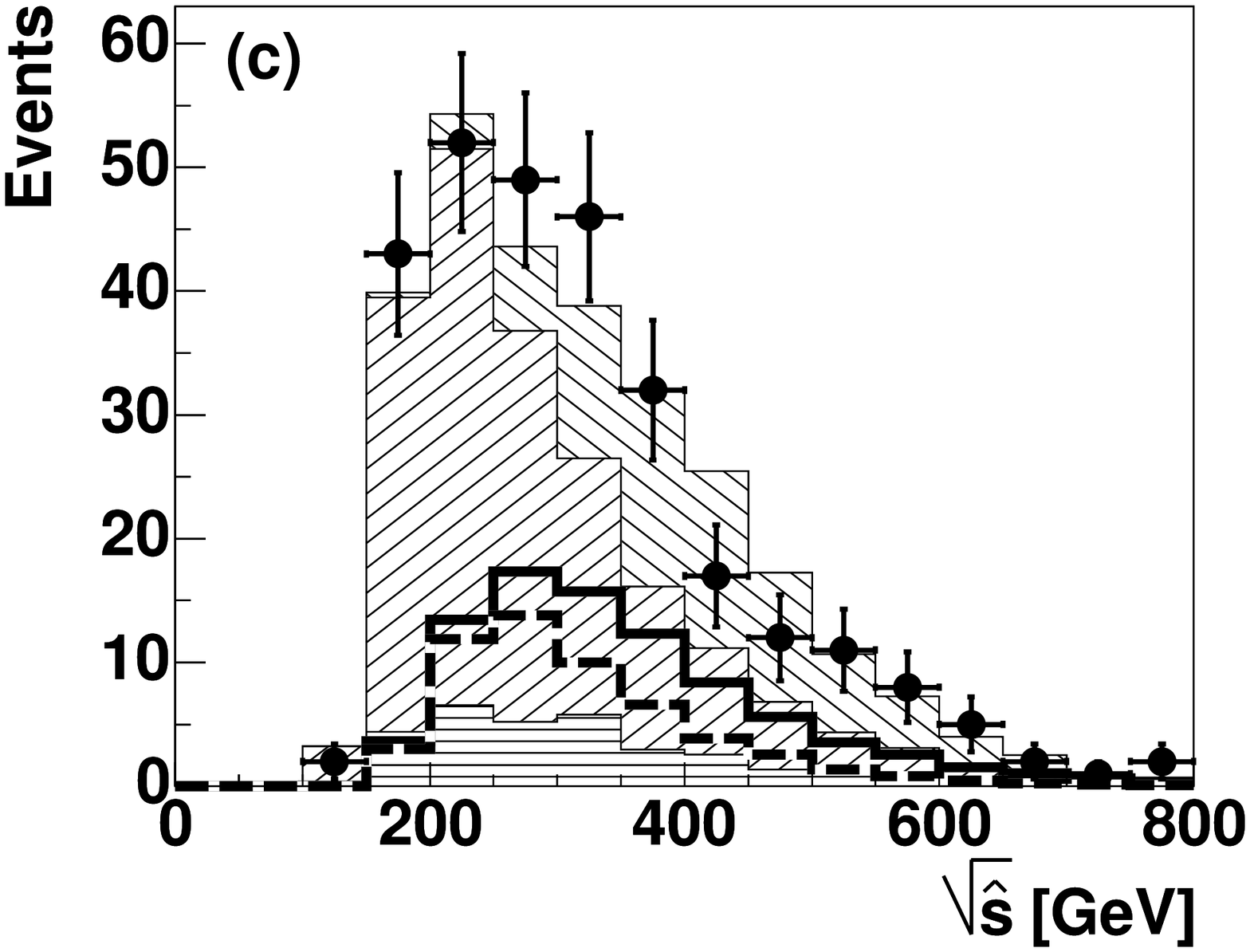}
\includegraphics[width=0.385\textwidth]
{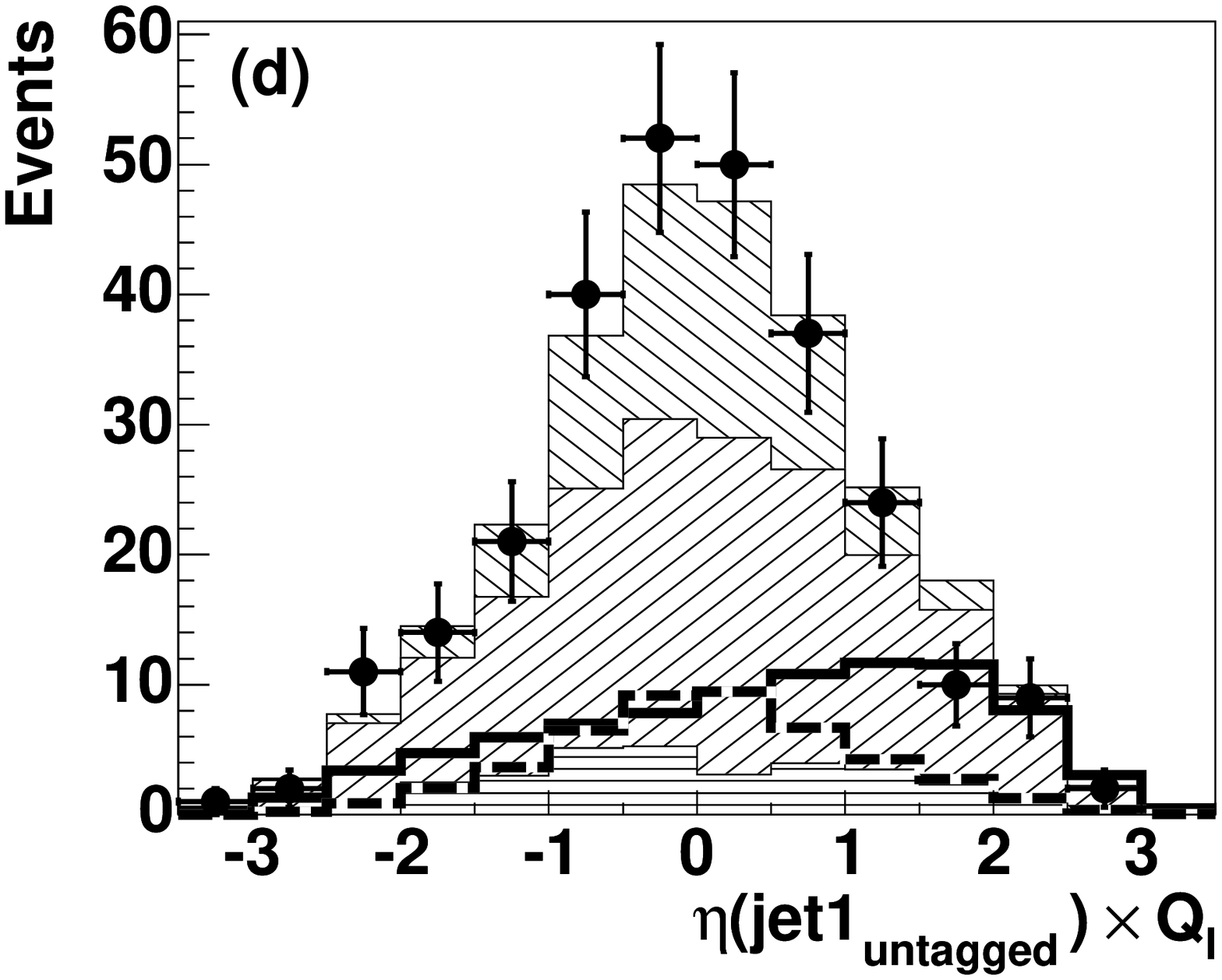}
\end{center}
\caption[var-yield-compare]{Comparison of signal, background, and 
data for the electron and muon
channels combined, requiring at least one tag, for four representative
neural network input variables.  Shown are 
(a) the transverse momentum of the leading untagged jet, 
(b) the invariant mass of the reconstructed top quark using the
leading tagged jet,
(c) the invariant mass of the final state system, 
and 
(d) the pseudorapidity of the leading untagged jet multiplied by the
charge of the lepton. 
Signals are multiplied by ten.}
\label{var-yield-compare}
\end{figure*}

Since the input variables do not depend on the lepton flavor,
the electron and muon analyses utilize the same variables.
However, owing to different lepton resolutions and pseudorapidity ranges,
we construct separate networks for them.
Therefore, four neural networks
are used for the signal-background pairs
($tb$-$W\bbbar$, $tb$-$\ttbar$, $tqb$-$W\bbbar$, $tqb$-$\ttbar$) for
each of the electron and muon channels.

Figure~\ref{var-yield-compare}
shows distributions of four representative variables.
We reconstruct the final state top quark from
the reconstructed $W$~boson and a jet as follows.
The $W$~boson is reconstructed from the isolated lepton and the missing
transverse energy. The $z$-component of the neutrino momentum ($p_{z}^{\nu}$) is
calculated using a $W$~boson mass constraint, choosing the solution
with smaller $|p^\nu_z|$ from the two possible solutions~\cite{Kane:1989vv}.
In the $s$-channel analysis, the
top quark is reconstructed from the $W$~boson and the ``best'' jet~\cite{d0runI}. 
The best jet is defined as the jet in each event
for which the invariant mass of the reconstructed $W$~boson
and the jet system is closest to $m_{t}=175$~GeV. 
In the $t$-channel analysis, the top quark is reconstructed from
the $W$~boson and the leading $b$-tagged jet.
Using these two methods we are able to correctly identify the $b$-quark jet from the
top quark decay in about 90\% of the signal events.

%
%
%
%
\begin{figure*}[h!tbp]  
\begin{center}
\includegraphics[width=0.385\textwidth]
{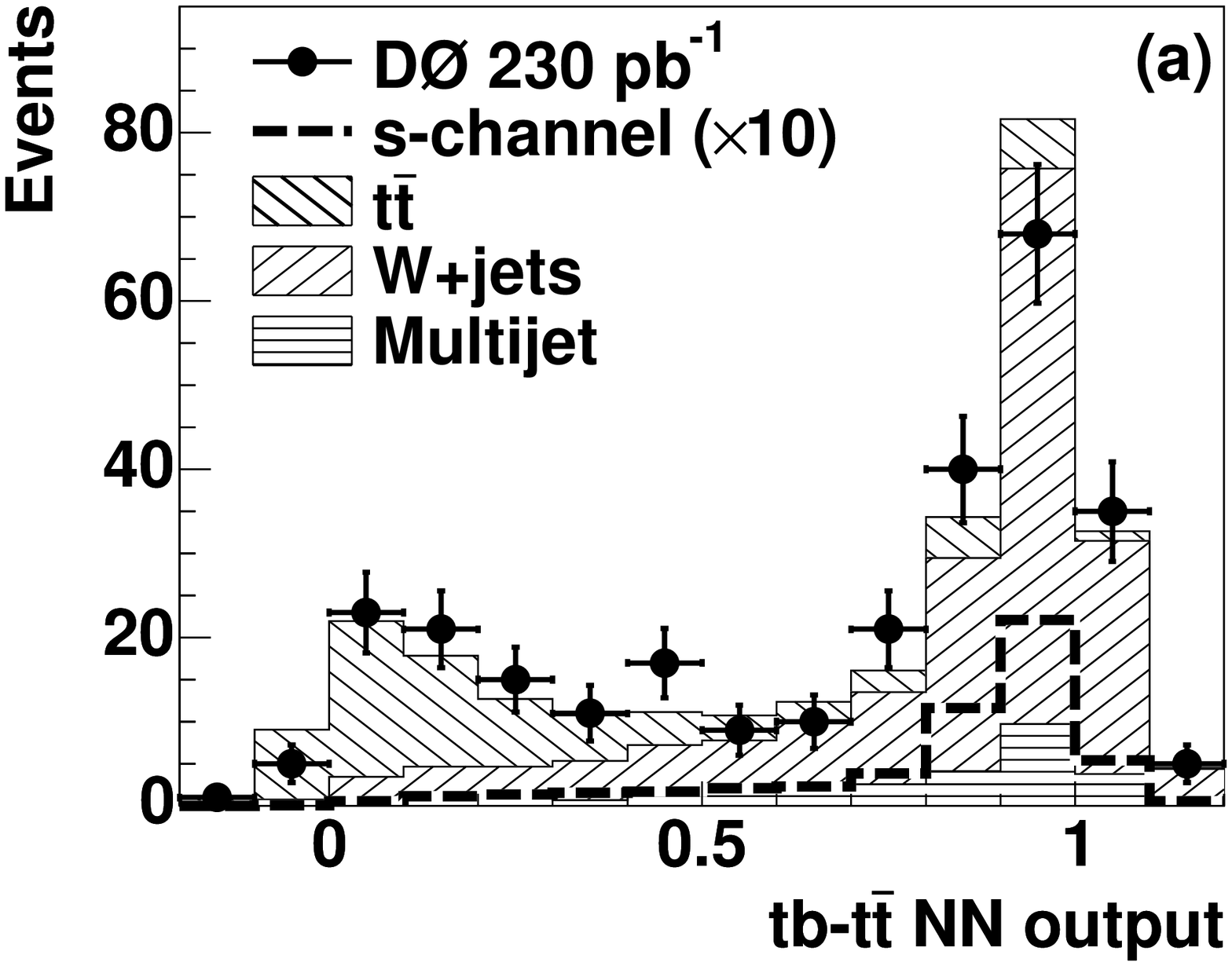}
\includegraphics[width=0.385\textwidth]
{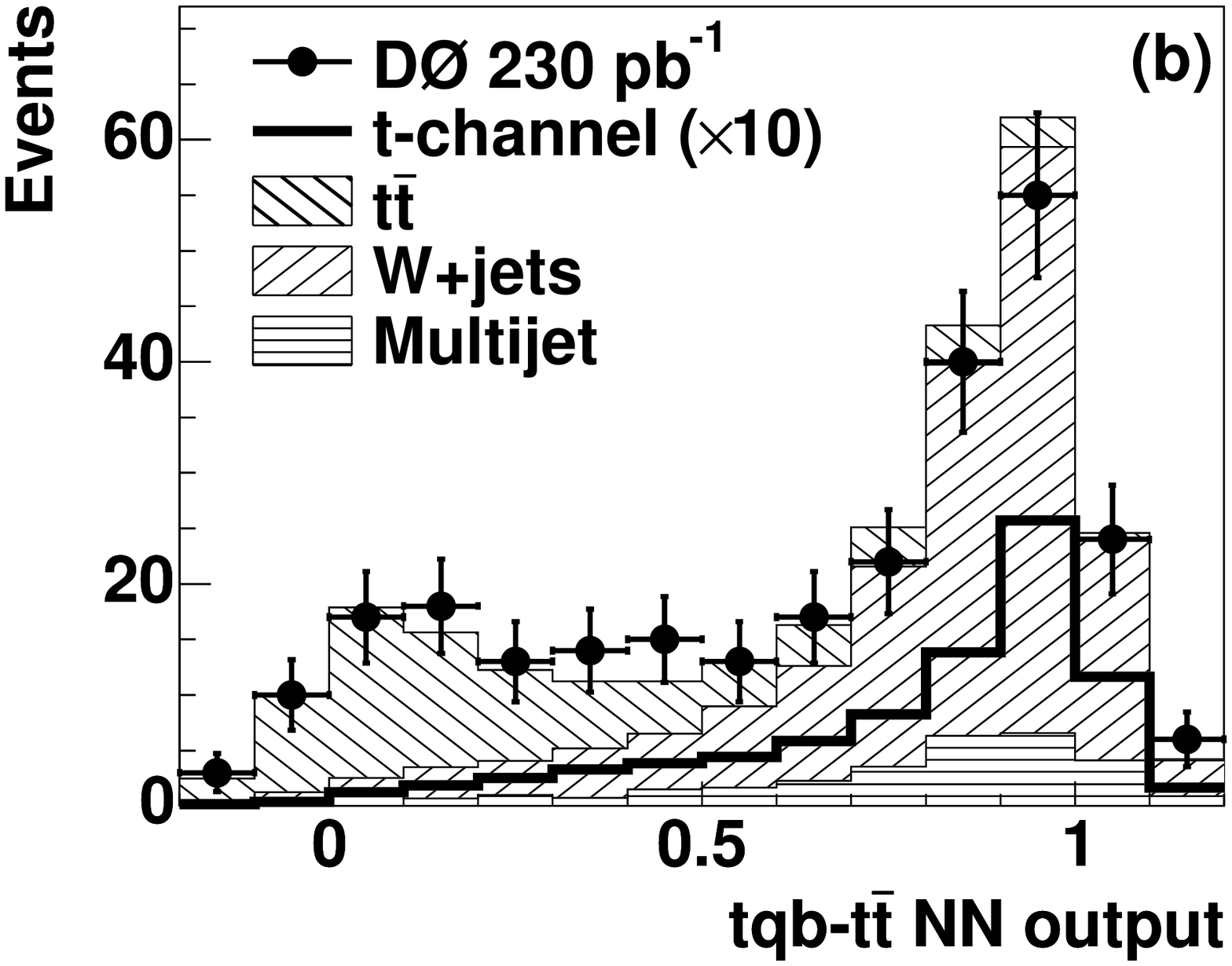}
\includegraphics[width=0.385\textwidth]
{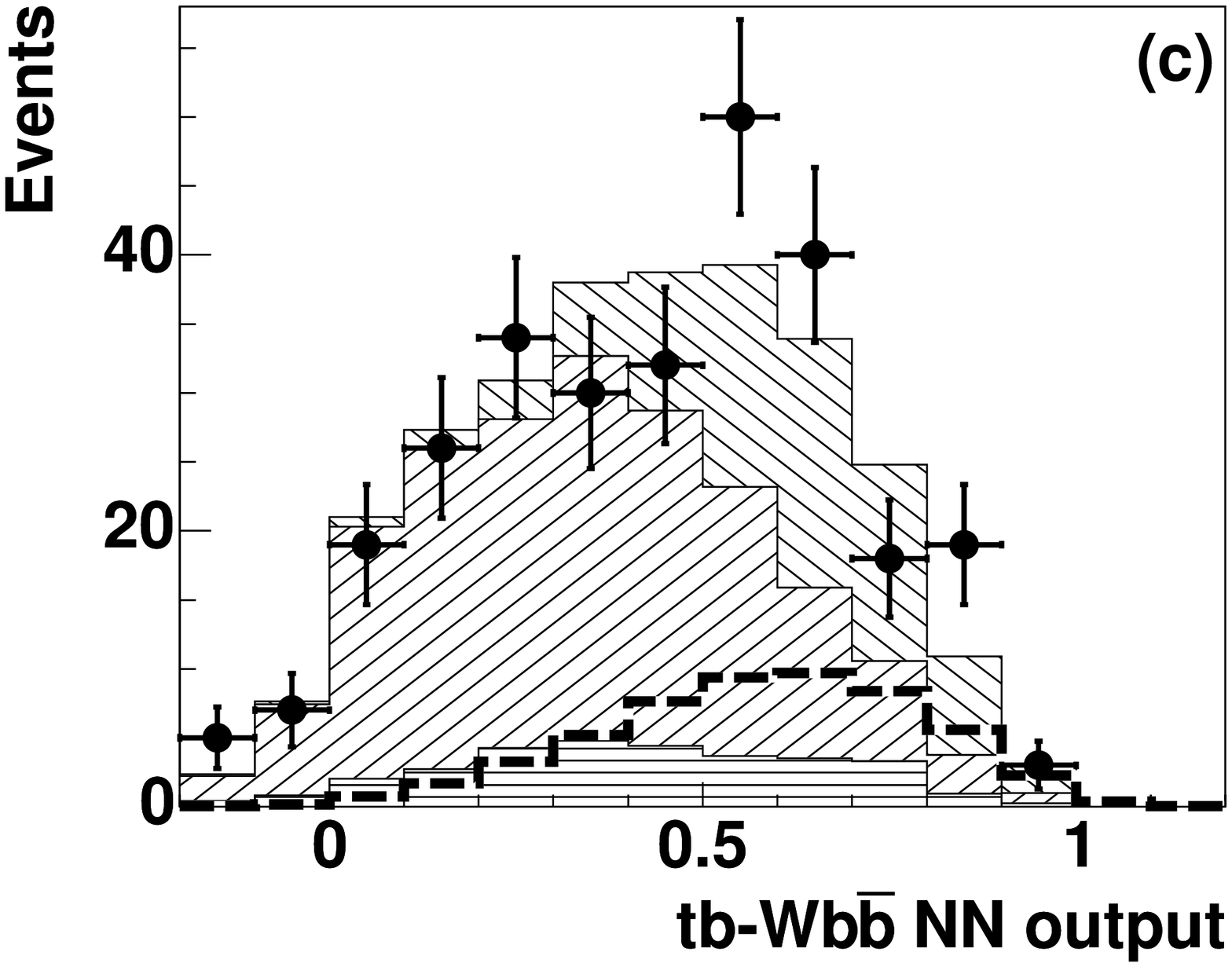}
\includegraphics[width=0.385\textwidth]
{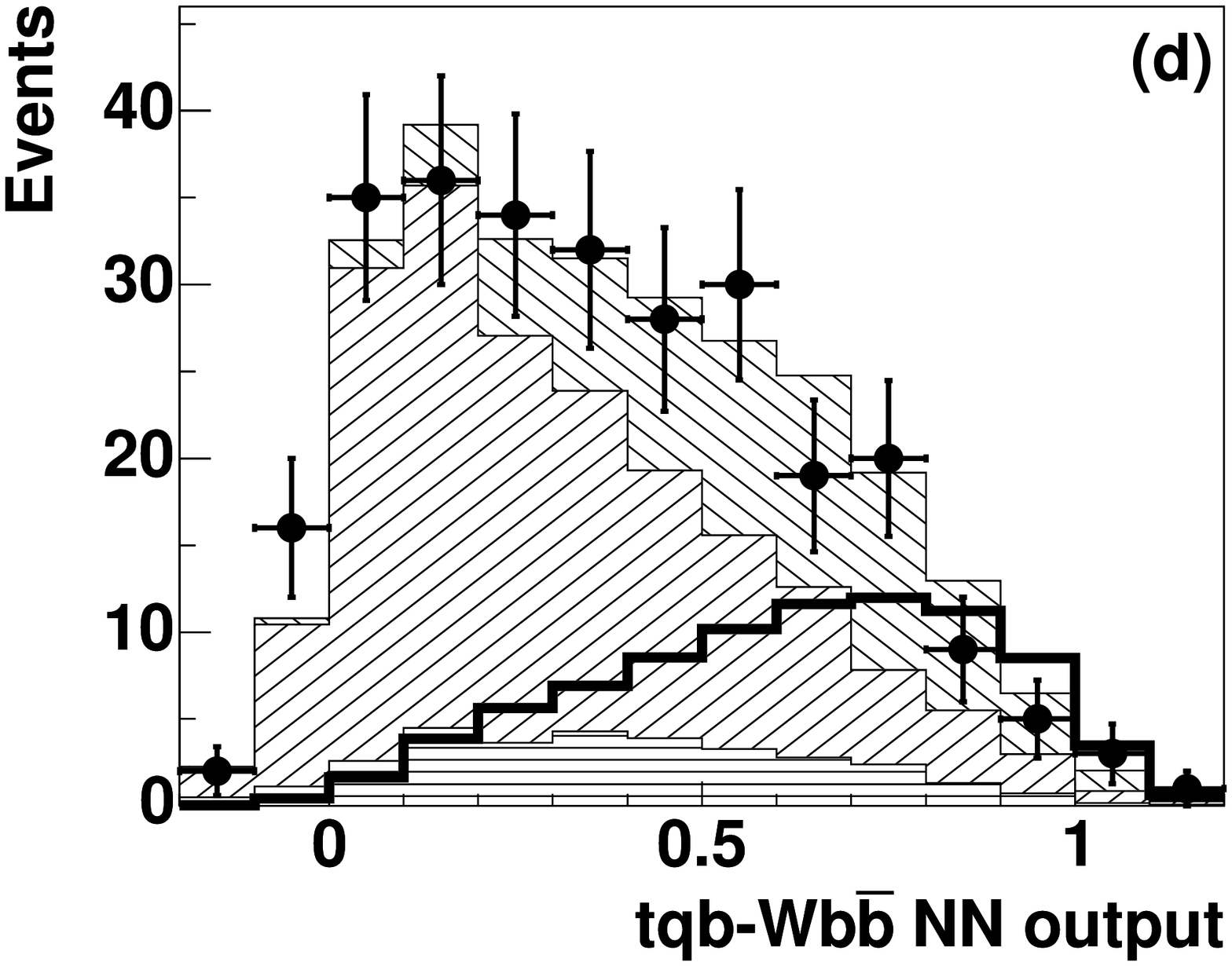}\\
\end{center}
\caption[nn-yield-compare]{Comparison of signal, background, and 
data for the neural network outputs, for the electron and muon
channels combined, requiring at least one tag.  This figure shows 
(a) the $tb$-$\ttbar$ filter, 
(b) the $tqb$-$\ttbar$ filter,
(c) the $tb$-$W\bbbar$ filter, and
(d) the $tqb$-$W\bbbar$ filter. Signals are multiplied by ten.}
\label{nn-yield-compare}
\end{figure*}
Figure~\ref{nn-yield-compare}
shows the outputs of the neural networks for
the data and the expected backgrounds, as well as signals for the electron and
muon channels combined. 
The neural network output in {\sc mlpfit} is around one for signal events
and around zero for background events, but it is not constrained to
the interval $[0,1]$. 
The $\ttbar$ networks separate signal and $\ttbar$ backgrounds
efficiently. The $Wb\bar{b}$ networks are less efficient for
the $W$+jets backgrounds because the event kinematics are similar
between signal and background.  
%
%
\section{Systematic uncertainties}
Systematic uncertainties are evaluated for the Monte Carlo signal
and background samples, separately for the electron and muon channels and for each 
$b$-tag multiplicity. 
The most important sources of systematic uncertainty are listed in
Table~\ref{tab:systematics}. 
The systematic uncertainty on the shapes of the distributions is also taken
into account for the contributions from $b$-tag modeling, jet energy calibration,
jet identification, and trigger modeling.
In order to evaluate the total uncertainty, we consider all sources of
systematic uncertainties for all samples and their correlations. 
The total uncertainty on the signal acceptance for single-tagged events is
13\% for the $s$-channel and 15\% for the $t$-channel, and for
double-tagged events it is 24\% for the $s$-channel and 28\% for the
$t$-channel. The total uncertainty on the background yield is 10\% for the
single-tagged samples and 26\% for the double-tagged samples.

%
%
\begin{table}[h!tbp]
\begin{center}
\caption{Range of systematic uncertainty values for the various Monte
Carlo signal and background samples in the different
analysis channels. 
}
\begin{ruledtabular}
\begin{tabular}{lc}
Source of               &  Uncertainty \\
systematic uncertainty  &  range (\%)\\
\hline
Signal and background acceptance \\
~~$b$-tag modeling      &  5 -- 20  \\ 
~~jet energy calibration&  1 -- 15  \\ 
~~trigger modeling      &  2 -- 7~  \\
~~jet fragmentation     &  5 -- 7~  \\
~~jet identification    &  1 -- 13  \\
~~lepton identification &   4  \\ 
Background normalization \\
~~theory cross sections &  2 -- 18  \\
~~$W$+jets flavor composition& 5 -- 16   \\
Luminosity              &   6.5 \\ 
\end{tabular}
\end{ruledtabular}
\label{tab:systematics}
\end{center}
\end{table}
%
%
\section{Cross section limits}
The observed data are consistent with the background predictions for 
all eight analysis channels. We therefore set upper limits on the 
single top quark production 
cross section separately in the $s$-channel and
$t$-channel searches using a Bayesian
approach~\cite{IainTM2000}. 
In each search, two-dimensional histograms are constructed from
the $W\bbbar$ vs. $\ttbar$ neural network outputs. 
A likelihood is built from these histograms for signal,
background, and data, as a product over all channels (electron and muon,
single and double tags) and bins.
We assume a Poisson distribution for the observed number of events in each bin and
a flat prior probability for the signal cross section. The prior for
the combined signal acceptance and background yields is a multivariate
Gaussian with uncertainties and correlations described by a covariance
matrix. 
Finally, we compute the posterior probability density as a function of
the production cross section. 

\begin{figure}[!h!tbp]
\begin{center}
\includegraphics[width=0.42\textwidth]
{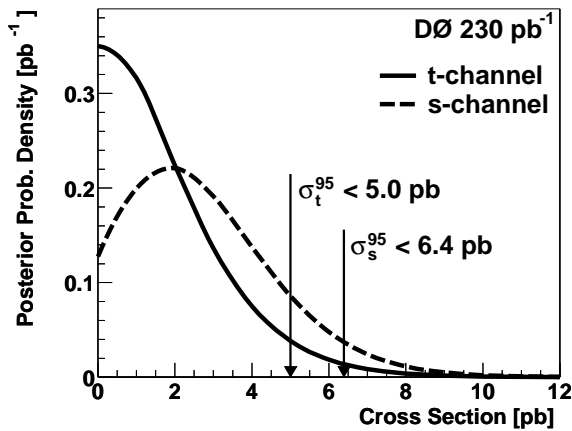}
\end{center}
\caption{The Bayesian posterior probability density as a function of the single top
quark cross section for the $s$-channel and $t$-channel searches.}
\label{nn-posterior-1d}
\end{figure}

%
%
The Bayesian posterior probability densities are shown in
Fig.~\ref{nn-posterior-1d} for both the
$s$-channel and $t$-channel searches. 
The corresponding upper limits at
the 95$\%$ C.L. are 6.4~pb in the $s$-channel and 5.0~pb in 
the $t$-channel. 
The sensitivity of these measurements is given by the expected upper
limits obtained by setting the observed number of events to the
background prediction in each bin. 
The expected upper limits are 4.5~pb in the $s$-channel search and 5.8~pb in the
$t$-channel search. 
%
%
\section{Conclusions}
No evidence is found for electroweak production of single top quarks
in 230~pb$^{-1}$ of data collected with the \dzero detector at
$\sqrt{s}=1.96$~TeV. 
The data consist of events in the electron and muon final states
with at least one $b$-tagged jet. 
We build binned likelihoods from the output of neural networks to
set upper limits at the 95\% C.L.
The measured $s$-channel limit is 6.4~pb and the measured $t$-channel limit is
5.0~pb. These upper limits represent significant improvements over previously published
results~\cite{d0runI,Acosta:2001un,RunII:cdf_result} due to the larger
data set as well as the use of a multivariate analysis technique together with
shape information from the resulting output distributions. 
They approach the region of sensitivity for models of
physics beyond the standard model, such as a fourth quark-generation
scenario or flavor-changing neutral-currents~\cite{Tait:2000sh}. 
%
%
\newpage
\section*{Acknowledgements}
%
We thank the staffs at Fermilab and collaborating institutions, 
and acknowledge support from the 
DOE and NSF (USA);
CEA and CNRS/IN2P3 (France);
FASI, Rosatom and RFBR (Russia);
CAPES, CNPq, FAPERJ, FAPESP and FUNDUNESP (Brazil);
DAE and DST (India);
Colciencias (Colombia);
CONACyT (Mexico);
KRF (Korea);
CONICET and UBACyT (Argentina);
FOM (The Netherlands);
PPARC (United Kingdom);
MSMT (Czech Republic);
CRC Program, CFI, NSERC and WestGrid Project (Canada);
BMBF and DFG (Germany);
SFI (Ireland);
Research Corporation,
Alexander von Humboldt Foundation,
and the Marie Curie Program.
%


\begin{thebibliography}{99}
\bibitem{fiducial_endnote}
Pseudorapidity is defined as $\eta = - \ln(\tan \frac{\theta}{2})$, 
where $\theta$ is the polar angle with the origin at the primary vertex. 
Detector fiducial regions are defined
by detector pseudorapidity  $\eta_{\rm det}$ which is calculated with
the origin at the nominal center of the detector ($z=0$).
\end{thebibliography}
\end{document}